# Divertor heat fluxes and profiles during mitigated and unmitigated Edge Localised Modes (ELMs) on the Mega Amp Spherical Tokamak (MAST)

**A.J. Thornton**[*], A. Kirk, I.T. Chapman, J.R. Harrison and the MAST Team

*EURATOM/CCFE Fusion Association, Culham Science Centre, Abingdon, Oxon, OX14 3DB,UK*

**Abstract**

Edge localised modes (ELMs) are a concern for future devices as they can limit the operational lifetime of the divertor. The mitigation of ELMs can be performed by the application of resonant magnetic perturbations (RMPs) which act to degrade the pressure gradient in the edge of the plasma. Investigations of the effect of RMPs on MAST have been performed in a range of plasmas using perturbations with toroidal mode numbers of n=3, 4 and 6. It has been seen that the RMPs increase the ELM frequency, which gives rise to a corresponding decrease in the ELM energy. The reduced ELM energy decreases the peak heat flux to the divertor, with a three fold reduction in the ELM energy, generating a 1.5 fold reduction in the peak heat flux. Measurements of the divertor heat flux profile show evidence of strike point splitting consistent with modelling using the vacuum code ERGOS.

# 1. Introduction

An edge localised mode (ELM) is a pressure driven instability which ejects a significant fraction of energy and particles from the confined plasma [1]. ELMs occur in the high confinement regime of tokamak operation, which is expected to be the baseline operating scenario for ITER [2]. The cyclical nature of the ELM heat flux has been shown to lead to melting and cracking of tungsten at relativity low heat fluxes (>1.6 MJ m$^{-2}$) which would greatly limit the operational lifetime of the ITER divertor [3]. Therefore, for ITER, a means of mitigating the heat loads from ELMs is essential. One such method of ELM mitigation is the use of resonant magnetic perturbations (RMPs) to degrade the edge pressure gradient and increase ELM frequency [4]. ELM mitigation via RMPs have been seen to be effective on several machines [4, 5, 6], and exploit the experimental observation that the product of the ELM frequency and the energy per ELM is constant. Hence, increasing the ELM frequency leads to a reduction in the peak divertor heat loads seen per ELM, in proportion to the change in the ELM wetted area.

# 2. ELM coils on MAST

MAST is equipped with a set of 18 RMP coils for ELM mitigation, with twelve coils equally spaced around the lower half of the machine and six coils equally spaced around the upper half of the machine [7]. The coils can be used to apply perturbations with various toroidal mode numbers (n = 3, 4, 6). The results reported in this paper use an n=3 perturbation in double null (DN) plasmas and n=4 and 6 perturbation in lower single null (LSN) plasmas. These RMP coil configurations are chosen to provide the optimal matching between the resonant component of the perturbation and the plasma q profile [4]. In addition, the higher mode number perturbations are used in LSN plasmas for ELM mitigation as significant core rotation braking is seen in these plasmas with the n=3 perturbations [8].

## 3. Divertor heat load measurements

The heat flux to the divertor on MAST can be monitored using infra-red (IR) thermography. The infra red thermography system consists of a medium wave IR (4.5-5 μm) camera and a long wave IR (7.9-9 μm) camera which monitor the divertor surfaces [9]. The long wave IR (LWIR) camera has been used to measure the heat flux to the lower divertor in LSN plasmas and the upper divertor in DN plasmas at frame rates of up to 14.5kHz. The medium wave IR (MWIR) camera has been used to measure the heat flux to the lower divertor in both DN and LSN plasmas at frame rates of up to 5kHz. The routine setup for the IR cameras gives a spatial resolution of the long and medium wave cameras of 7.5 and 5 mm respectively at the target which is equivalent to 1.5 and 1mm at the outboard midplane.

A key issue for IR thermography is the effect of surface layers on the measured heat fluxes. The surface layer is accounted for via the use of a surface layer coefficient, $\alpha$, in the THEODOR heat transport code [10]. In order to determine the surface layer coefficient for the ELMs, energy balance has been performed by comparing the IR measured energy to the divertor and the calculated energy loss from the ELM determined by EFIT reconstruction [11]. The surface layer coefficient is then adjusted to give energy balance between these two quantities. The energy balance for LSN discharges where both the LWIR and MWIR cameras are viewing the same divertor is shown in Figure 1. It can be seen from Figure 1 that energy balance has been achieved in these discharges using $\alpha_{LWIR}^{LOWER} = 120$ kW K$^{-1}$ m$^{-2}$ and $\alpha_{MWIR}^{LOWER} = 50$ kW K$^{-1}$ m$^{-2}$, which are consistent with those required for L mode energy balance. The difference in $\alpha$ between the two cameras is due to the effect of hot spots on the surface layer coefficient in MAST as previously studied by Delchambre et al [12]. Confirmation of the surface layer coefficients is made by cross calibration between the MWIR and LWIR cameras. Measurements of the peak heat flux recorded on the MWIR and LWIR cameras observing the same toroidal location of the divertor in LSN discharges, show that the peak recorded heat flux is the same for both cameras, confirming the choice of surface layer coefficient. Similar analysis on DN plasmas gives the surface layer coefficient for the LWIR camera on the upper divertor $\alpha_{LWIR}^{UPPER} = 60$ kW K$^{-1}$ m$^{-2}$.

## 4. Effect of ELM mitigation

The ELM frequency with and without RMPs can be found for a range of discharges and is shown in Figure 2 plotted against the energy loss per ELM. The application of RMPs increases the ELM frequency, with a corresponding decrease in the ELM energy. A typical mitigated discharge shows a four fold increase in ELM frequency, with an equivalent decrease in the ELM energy. The mitigated ELMs follow the trend of the natural ELMs, showing that the mitigated ELMs behave as small natural ELMs and that the product of the ELM frequency and the ELM energy loss is a constant.

The increased ELM frequency and corresponding decrease in the energy loss per ELM give rise to lower divertor heat loads in the case of the mitigated ELMs, as shown in Figure 3. The mitigated and natural ELMs follow the same trend which shows that decreasing the ELM size by three gives rise to the peak heat flux decreasing by a factor of 1.5 over the natural ELMs. The decrease in the peak heat flux during mitigation is accompanied by a decrease in the ELM wetted area, as shown in Figure 4. The ELM wetted area is determined for the outer strike point using the definition in Jachmich et al [13]. The wetted area in natural ELMs is approximately 0.6 m$^2$ for the largest ELMs (15kJ), decreasing to 0.4 m$^2$ when the ELM size decreases by a factor of three. The decrease in the ELM wetted area is consistent with the trend seen in the peak heat flux reduction and is seen on other machines [13]. The peak heat flux to the divertor, q, can be approximated as $\Delta E / A_{ELM} t_{ELM}$, where $\Delta E$ the energy loss per ELM and $t_{ELM}$ is the duration of the ELM heat flux which is seen to be constant for natural and mitigated ELMs. The data shows that the reduction in peak heat flux for a given reduction in ELM energy scales consistently with this formula, when the change in the area is taken into account.

The energy balance to the inner (high field side) divertor and the outer (low field side) divertor in LSN plasmas is shown in Figure 5. The ratio of the energy to the high field side, $E_{HFS}$, to the low field side, $E_{LFS}$, is seen to be 0.3 in MAST both with and without the application of RMPs. The ratio $E_{HFS}/E_{LFS}$ is seen to be significantly different on other machines, where the energy to the inner divertor is 1.5 times that to the outer divertor [14].

**5. Strike point splitting in the presence of RMPs**

Experiments performed on MAST have seen splitting in L mode plasmas [15] which is accompanied with density pump out from the plasma [16]. The L mode splitting is seen to be well aligned to the expected location of the splitting as predicted by the vacuum modelling code ERGOS [17].

Observations of the poloidal plasma cross section have shown lobe structures extending to the divertor surface, which are consistent with vacuum modelling [7]. Imaging of the divertor with CII (515nm) filtering and a spatial resolution of 2mm have been used to measure the splitting of the particle flux during inter-ELM periods. The divertor CII emission profile for an inter-ELM period in a LSN discharge is shown in Figure 6. The splitting can be clearly seen inter-ELM on the filtered imaging and shows agreement in location to that predicted by vacuum modelling. The splitting cannot be seen on the IR cameras during the inter-ELM periods in the H mode plasmas. The plasma rotation acts to screen the applied perturbation in H mode plasmas. Screening affects the length of a given lobe in the divertor footprint, shortening lobes compared to the unscreened case [15]. Hence, the lack of splitting can be explained by the lobes not extending to the toroidal angle of the IR camera. Alternatively, there may not be enough energy deposited into the ELM footprint during the inter-ELM phase. The low energy prevents the splitting from being observed except during the ELMs where the energy to the target is increased, as seen on DIII-D [18].

During ELMs, previous studies [19, 20] have shown that strike point splitting can be generated due to the filamentary nature of the ELM. It would be expected that the splitting from the ELM filaments would have a random structure, which would not persist if profiles from several ELMs are averaged together. The IR profiles in a set of ELMs with RMPs applied can be coherently averaged to determine if the splitting can be seen during the ELM heat pulse on MAST. Figure 7 shows the averaged IR profiles for three different periods during the ELM cycle, defined to start with the rise in $D_\alpha$ at $t_0$. At the peak of the ELM $D_\alpha$ emission ($t_0 + 100$ μs), the heat flux is seen to be uniform, exhibiting no splitting. During the increase in the heat flux ($t_0 + 150$ μs) a coherent structure is seen on the IR profile, which is consistent with splitting of the strike point. The structure decreases in amplitude in the later stages of the ELM cycle ($t_0 + 200$ μs), which could be due to the filamentary structure of the ELMs washing out the splitting as a result of the separated filaments arriving in random locations at the divertor.

## 6. Summary

ELM mitigation via RMP has been seen to be effective on MAST at decreasing the peak heat flux to the divertor by 1.5 times for a three fold reduction in the energy loss per ELM. Measurements of the particle and heat fluxes to the divertor during ELM mitigation has shown evidence for strike point splitting inter-ELM in particle flux and during the initial phase of the ELM cycle for the heat flux. The location of the strike point splitting is consistent with modelling of the divertor strike point determined using the vacuum modelling code ERGOS. The lack of splitting in the heat flux profiles during the inter-ELM period has been seen on other machines. However, as the splitting is seen in the particle flux inter-ELM using CII imaging, it would be expected that the splitting should also be present in the heat flux. Further measurements of the heat flux to the divertor using increased spatial resolution and longer integration times are required to confirm this result.

**Acknowledgements**

This work was part-funded by the RCUK Energy Programme under grant EP/I501045 and the European Communities under the contract of Association between EURATOM and CCFE. The views and opinions expressed herein do not necessarily reflect those of the European Commission.

**Figure captions**

Figure 1. a) Energy balance in LSN discharges for the MWIR (black circles) and LWIR camera (red triangles) for both cameras viewing the same divertor simultaneously.

Figure 2. The energy loss per ELM ($\Delta W_{ELM}^{EFIT}$) as determined using EFIT reconstruction plotted as a function of the ELM frequency for natural (black circles) and mitigated (red triangles) ELMs.

Figure 3. The peak heat flux to the outer divertor as a function of the energy loss per ELM ($\Delta W_{ELM}^{EFIT}$) for natural (black circles) and mitigated (red triangles) ELMs.

Figure 4. The ELM wetted area on the outer divertor as a function of the energy loss per ELM ($\Delta W_{ELM}^{EFIT}$) for natural (black circles) and mitigated (red triangles) ELMs.

Figure 5. The energy load the inner strike point from IR thermography as a function of the energy to the outer strike point. The energy to the inner divertor is three times lower than the energy to the outer divertor.

Figure 6. Imaging of the divertor in CII (515 nm) light can be used to asses the splitting of the strike point in particle flux. The black trace shows the emission profile during the coils on phase and the red trace shows the emission during the coils off phase. The amplitude of the emission is normalised to the peak of each profile. The blue trace shows the predicted location of the splitting from vacuum modelling.

Figure 7. Coherently averaged IR heat flux profiles for three periods during the ELM cycle. The onset of the ELM, $t_0$, is defined by the rise in the $D_\alpha$ emission at the midplane. The first profile is taken at $t_0+100$ μs and shows no splitting during the inter-ELM period. The red trace shows the heat flux profile at $t_0+150$ μs, which corresponds to the rising phase of the divertor heat flux. The blue profile is taken at $t_0+200$ μs which correspond to the time where the divertor heat flux reaches a maximum.

**Figures**

**Figure 1**

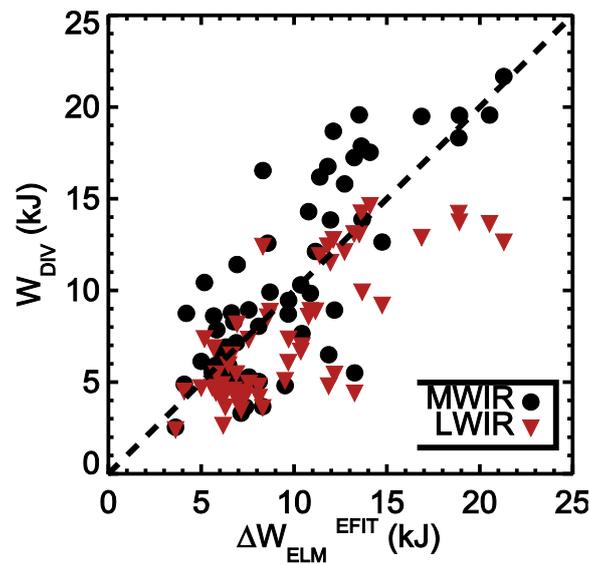

**Figure 2**

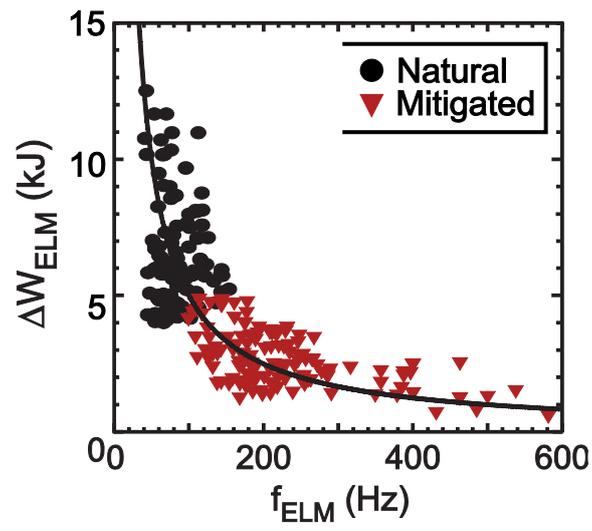

**Figure 3**

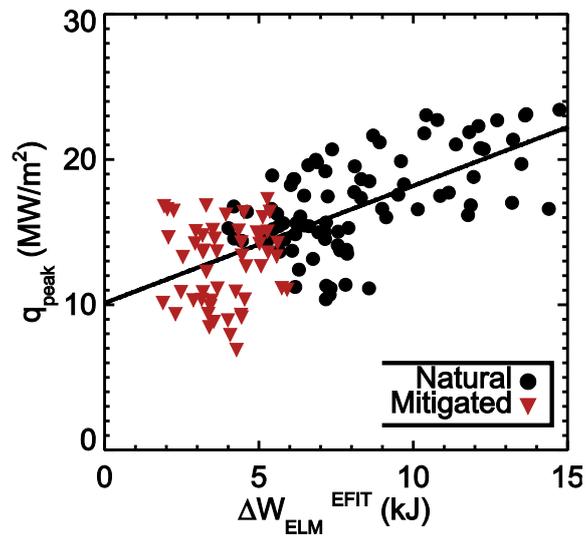

**Figure 4**

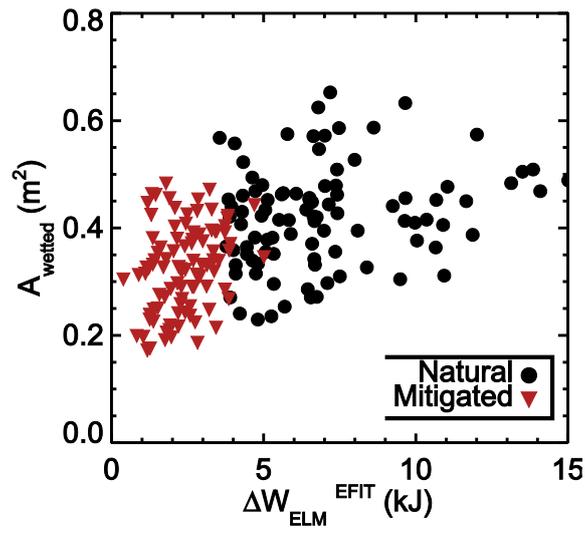

**Figure 5**

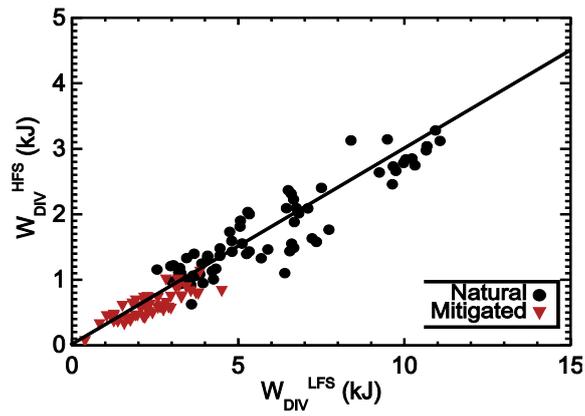

**Figure 6**

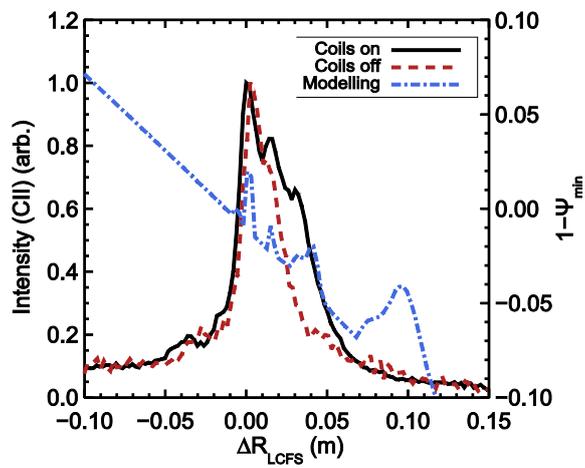

**Figure 7**

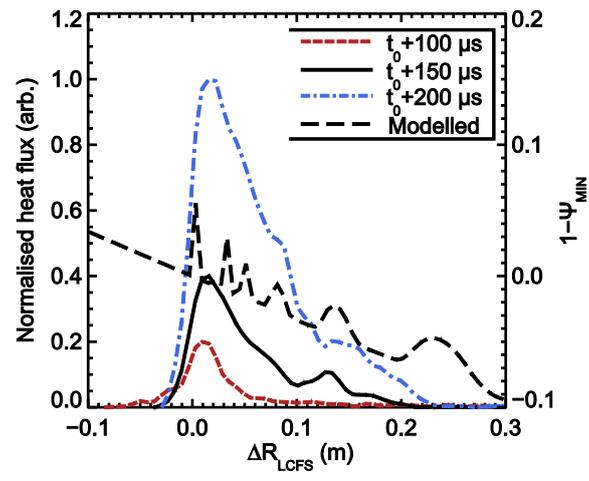